\documentclass{elsarticle}

\usepackage{amsmath}
\usepackage{amsfonts}
\usepackage{amssymb}

\usepackage{tikz}
\usetikzlibrary{positioning}
\usetikzlibrary{arrows.meta, decorations.pathreplacing, shadows}
\usepackage{color}
\usepackage[utf8]{inputenc} 
\usepackage[T1]{fontenc}

\begin{document}

\begin{frontmatter}

\title{A deep learning approach to predict significant wave height using long short-term memory}

\author[inst1]{Felipe C. Minuzzi}
\ead{felipe.minuzzi@ufrgs.br}
\author[inst1,inst2]{Leandro Farina}
\ead{farina@alum.mit.edu}

\affiliation[inst1]{organization={Institute of Mathematics and Statistics, Federal University of Rio Grande do Sul (UFRGS)},addressline={Av. Bento Goncalves 9500},city={Porto Alegre},postcode={PO Box 15080}, state={RS},country={Brazil}} 
\affiliation[inst2]{organization={Center for the Study of Coastal and Oceanic Geology (CECO), Federal University of Rio Grande do Sul (UFRGS)},addressline={Av. Bento Goncalves 9500},city={Porto Alegre},postcode={Pr\'{e}dio 43.125}, state={RS},country={Brazil}}

\begin{abstract}
 We present a framework for forecasting significant wave height on the Southwestern Atlantic Ocean
 using the long short-term memory algorithm (LSTM), trained with the ERA5 database available through Copernicus Climate Data Store (CDS) implemented by ECMWF (European Center for Medium Range Forecast) and also with buoy data. The predictions are made for seven different locations in the Brazilian coast, where buoy data are available, ranging from shallow to deep water. Experiments are conducted using exclusively  historical series at the selected locations and the influence of other variables as inputs for training is investigated. The results shows that a data-driven methodology can be used as a surrogate to the computational expensive physical models, with the best accuracy near $95\%$, compared to reanalysis data and and $87\%$ compared to buoy data. 
\end{abstract}



\end{frontmatter}

\section{Introduction}\label{intro}
Accurate predictions of ocean waves are essential for a handful of industries that define their operations based on the outcome of this type of analysis. Almost all engineering applications concerning the ocean, from navigation to renewable energy, going through offshore platforms, alerts of catastrophic events and geosciences research \cite{komen1996dynamics, cavaleri2007wave, ardhuin2019observing}, benefit from an accurate description of the sea state, of which wave heights are probably the most important parameter.

Is no novelty that ocean waves can be simulated with mathematical-physical models, and several state-of-the-art programs are available for such purpose (see, e.g., the WAVEWATCH III \cite{wavewatch} and SWAN \cite{booij1997swan}). However, powerful artificial intelligence algorithms are gaining visibility and  popularity with the increasing  public computational libraries available making it possible the analysis of large amount of data for several applications that are based on historical information. As several reanalysis reliable databases are available and continue to expand the time period for which they provide historical data, one important question has been posed: can data-driven models, with help of the artificial intelligence,  act as a physical model surrogate, with computational time and accuracy that are superior to the latter? 

The use of artificial intelligence, specially in the context of artificial neural networks (ANNs) in wave modelling has already been investigated and studied. In a pioneering research, Deo and Naidu \cite{deo1998real} used feed-forward networks with three different training approaches to forecast wave heights along the east coast of India. They used buoy observations data as input to the model, collected every three hour, from May 1983 until August 1984, resulting in 16 months of historical data. Different lead times were analysed and the predictions yielded accurate results. From this point on, ANNs have been used in several works for wave predictions. 

Short-term forecasts using ANNs trained with data from two sites offshore the Atlantic and the Irish Sea coasts of Ireland is reported in Makarynskyy \cite{makarynskyy2004improving}, and hourly forecasts of significant wave height and zero-up-crossing wave periods are made for $1-24$h time intervals. In another ANNs work, Makarynskyy and collaborators \cite{makarynskyy2005artificial} have used buoy data from Portugal's West coast, also to forecast significant wave height and zero-up-crossing wave period for 3, 6, 12 and 24h intervals. In one of the approaches, each parameter over every time interval is forecasted by a separate ANN, while in the second approach, only two ANNs are used to concurrently simulate the variables. Browne \textit{et al.} gave a comprehensive explanation of using ANNs to estimate waves near-shore and compared results with SWAN model. Results show that ANNs outperformed the physical model in simulations for seventeen near-shore locations around the continent of Australia over a 7 month period.

Agrawal and Deo \cite{agrawal2004wave} applied ANNs as an alternative to find interrelationships among certain characteristic wave parameters. Networks were trained for locations at the east coast of India and developed in order to estimate values of average zero-cross wave period, peak-spectral period, maximum spectral energy density and maximum wave height from the given value of significant wave height and also to evaluate the spectral width parameter from the spectral narrowness parameter. Krasnopolsky \textit{et al.} \cite{krasnopolsky2002neural} proposed an alternative to the complex mathematical formulations involved in forecast systems by approximating solutions of exact physical models using ANNs. They consider the UNESCO equation of state of the seawater (density of the seawater) and an approximation for the nonlinear wave–wave interaction. The non-linear interactions in wind wave spectra is also investigated by Tolman \textit{et al.} \cite{tolman2005neural} using ANNs, while Zamani \textit{et al.} \cite{zamani2008learning} forecasted significant wave heights for several hours ahead using buoy measurements in Caspian Sea using models based on ANNs.

Londhe and Panchang \cite{londhe2006one} applied ANNs within the feed-forward back propagation algorithm for four different lead times (6, 12, 18 and 24) to forecast significant wave heights in six different buoy locations at the Gulfs of Mexico, Alaska and Maine. Six network architectures were tested for each buoy location, and predictions were made for the period of 8 January until 31 December, 2004 for all buoys except one, for which the predictions ended in 16 September of that year. An accuracy of $86\%$ is obtained for the 6h lead time and between  $67\%$ and $83\%$ for 12h. Besides that point, as expected, the accuracy drops to $55\% - 71\%$ for 18h lead time and less than $63\%$ for the 24h lead time forecast. Despite the good results, a limitation highlighted by using ANNs in ocean wave predictions is the considerable under-predictions in the highest peaks, which can be explained by the dominance of smaller wave heights records in the datasets used to train the network. In an attempt to overcome this drawback, data from 2004 were added to the training phase (which had records of the Hurricane Ivan), and a tendency to catch the higher peaks was observed.

More recently, Campos \textit{et al.} \cite{campos2020improving, campos2017impr} developed a post-processing algorithm to improve ensemble averaging, as a replacement to the typical arithmetic ensemble mean, using neural networks trained with altimeter data. Similar techniques using nonlinear ensemble averaging, also based on neural networks, were studied in the Gulf of Mexico \cite{campos2019nonlinear}. James \textit{et al.} \cite{james2018machine} used supervised machine learning algorithms to estimate ocean-wave conditions at the Monterey Bay, California, to act as a surrogate to physical models, while O'Donncha \cite{o2018integrated} \textit{et al.} combined both ensemble physical model simulations and machine learning to derive an integrated technique that investigate uncertainty from modelling and generate a forecast that is better than the best individual model prediction. 

Not only ANNs were used to ocean wave predictions in the context of data-driven results, but also several other artificial intelligence techniques, such as support vector machines~\cite{browne2007near}, Bayesian optimization~\cite{cornejo2018bayesian}, genetic programming~\cite{nitsure2012wave,gaur2008real} and wavelets~\cite{oh2018real,prahlada2015forecasting}. Furthermore, Deep learning, i.e., neural networks within a deep range that uses great amount of data, has already been shown to be a powerful technique for oceanography predictions~\cite{zheng2020purely, choi2020real}.

The long short-term memory (LSTM) recurrent neural network was introduced by Hochreiter and Schmidhuber \cite{hochreiter1997long}, with the ability of learning long-term dependencies. This benefit gives the possibility to create a neural network that can use information of a long past to build its predictions. In a recent work, Pirhooshyaran and Snyder \cite{pirhooshyaran2020forecasting} used LSTM together with a sequence-to-sequence neural network to forecast and hindcast ocean waves, as well as to reconstruct missing data. Besides, feature selection has been employed based on nearby buoys data.

In the context of ocean wave modelling, given the comprehensive amount of significant wave height time series documented and available, LSTM seems like a good alternative to improve predictions of this variable specially due to its 'forget gate' layer, which gives the net the capability to remember long-term propagation of data, or forget them, when is suitable. Thus, the present work aims to investigate LSTM neural networks to forecast significant wave height on several locations in Brazil's coast, for four different lead times. We consider only the historic wave data to feed the net since, as it will be shown, additional features show no improvement in the final results. Our approach differs from the work in \cite{pirhooshyaran2020forecasting} both in methodology and goals. We choose to perform forecast with verification times within a time period of one month (or $744$ time-steps) and the training phase of the LSTM considers four different lead times, namely, $6$, $12$, $18$ and $24$. Moreover, we solely focus on LSTM predictions for significant wave height, using i) only this variable for uni-variate forecasts; ii) closely correlated variables, such as peak wave period and $10$m wind speed and iii) the best four variables based on Pearson's correlation coefficient, for multi-variate forecast. Training of the LSTM network is performed using both ERA5 reanalysis data, implemented by the European Center for Medium Range Forecast (ECMWF) \cite{copernicus, hersbach2020era5} and real buoy measurements. The former has a better accuracy in the prediction period when compared to its own training database, but the latter shows higher accuracy with respect to observations. As a contribution, we show an architecture for LSTM network that can be suitable for short-term forecasts of significant wave height, and potentially for other ocean waves variables as well, with a large decrease in computational time if compared to traditional physical models, and with accuracy, superior  to the reanalysis, when the training is done base exclusively with observational data.

This paper is structured as follows: In Section \ref{ml}, we show the mathematical and theoretical background of the machine learning algorithm used in this work, the LSTM . In Section \ref{data}, the data used is described and in Section \ref{method} the methodology of our framework is presented. Section \ref{results} shows the predictions results based on LSTM and comparisons with ERA5 and buoy observations. Finally, the conclusions are presented in Section \ref{conclusion}.

\section{Machine Learning} \label{ml}
Machine learning algorithms are capable of producing data-driven decisions, i.e, given the information, the computer learns from it and delivers solutions. Supervised learning is a type of machine learning that uses labelled data to train the algorithm, where both input and output data are given in this training phase. The algorithm learns the mapping function from the input to output variables, and after that, with a new set of data, the algorithm produces an output. 

Amongst machine learning techniques, artificial neural networks (ANN) can be seen as a machine designed to model the way in which the brain performs a particular task or function of interest \cite{haykin2009neural}. From a mathematical standpoint, they can be considered as multiple nonlinear regression methods able to capture hidden complex nonlinear relationships between input and output variables \cite{peres2015significant}.  

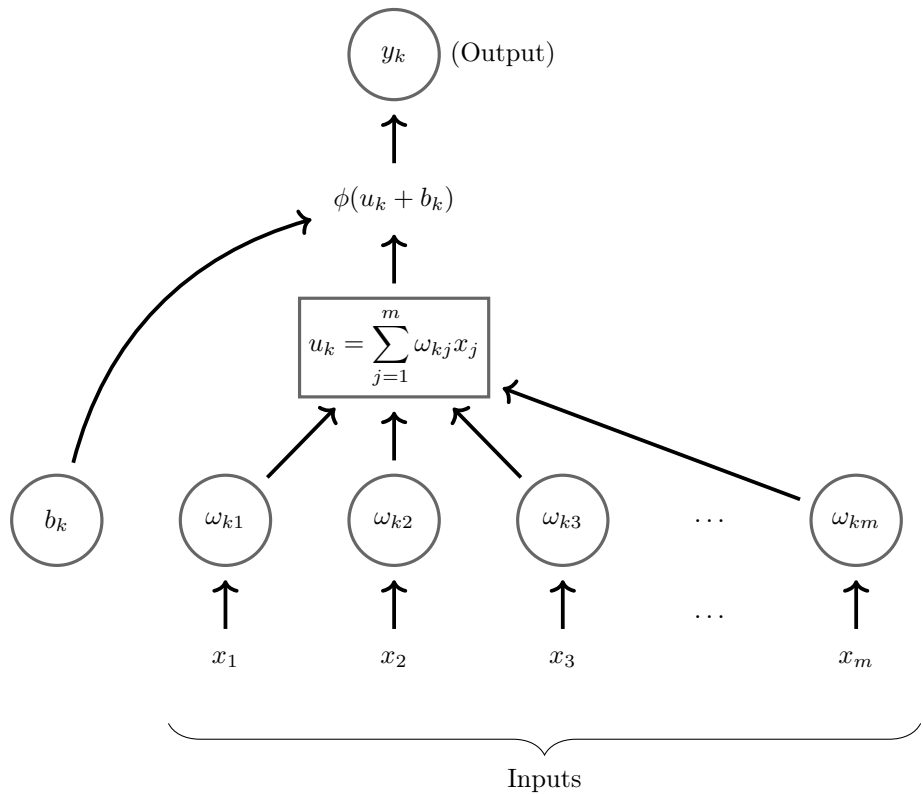
\begin{figure}
\begin{center}
\begin{tikzpicture}[roundnode/.style={circle, draw=black!60, fill=white!5, very thick, minimum size=12mm}, squarednode/.style={rectangle, draw=black!60, fill=white!5, very thick, minimum size=5mm}, node distance = 1cm]
    
    \node[squarednode]      (maintopic)     {$u_k = \displaystyle\sum_{j=1}^m \omega_{kj}x_j$};
    \node[roundnode]        (circle1)       [below  = of maintopic] {$\omega_{k2}$};
    \node[roundnode]        (belowcircle)   [left = of  circle1] {$\omega_{k1}$};
    \node[roundnode]        (bias)          [left = of  belowcircle] {$b_k$};
    \node[roundnode]        (circle2)       [right= of circle1] {$\omega_{k3}$};
    \node[]                 (circle3)       [right= of circle2] {$\ldots$};
    \node[roundnode]        (circle4)       [right= of circle3] {$\omega_{km}$};
    \node[]                 (circle5)       [above = of maintopic] {$\phi(u_k + b_k)$};
    \node[roundnode]        (circle6)       [label=right: (Output)] [above= of circle5] {$y_k$};
    \node[]                 (circle7)       [below= of belowcircle] {$x_1$};
    \node[]                 (circle8)       [below= of circle1] {$x_2$};
    \node[]                 (circle9)       [below= of circle2] {$x_3$};
    \node[]                 (circle10)      [below= of circle3] {$\ldots$};
    \node[]                 (circle11)      [below= of circle4] {$x_m$};
    
    \draw[->, shorten >= 5pt, shorten <= 5pt, line width=0.5mm] (belowcircle) edge (maintopic);
    \draw[->, shorten >= 5pt, shorten <= 5pt, line width=0.5mm] (circle1) edge (maintopic);
    \draw[->, shorten >= 5pt, shorten <= 5pt, line width=0.5mm] (circle2) edge (maintopic);
    \draw[->, shorten >= 5pt, shorten <= 5pt, line width=0.5mm] (circle4) edge (maintopic);
    \draw[->, shorten >= 5pt, shorten <= 5pt, line width=0.5mm] (maintopic) edge (circle5);
    \draw[->, shorten >= 5pt, shorten <= 5pt, line width=0.5mm] (bias) edge[bend left] (circle5);
    \draw[->, shorten >= 5pt, shorten <= 5pt, line width=0.5mm] (circle5) edge (circle6);
    \draw[->, shorten >= 5pt, shorten <= 5pt, line width=0.5mm] (circle7) edge (belowcircle);
    \draw[->, shorten >= 5pt, shorten <= 5pt, line width=0.5mm] (circle8) edge (circle1);
    \draw[->, shorten >= 5pt, shorten <= 5pt, line width=0.5mm] (circle9) edge (circle2);
    \draw[->, shorten >= 5pt, shorten <= 5pt, line width=0.5mm] (circle11) edge (circle4);
    \draw [decorate, decoration={brace,mirror,amplitude=12pt}] (-3,-5) -- (7,-5) node [midway,yshift=-0.3in] {Inputs};
    
\end{tikzpicture}
\end{center}
\caption{Schematic of a neuron in an artificial neural network.}
\label{fig1}
\end{figure}

In its simplest form, the structure of an ANN is based on a unit, or neuron ($u_k$), which receives a linear combination of weighted input and bias, i.e \cite{makarynskyy2004improving, haykin2009neural},
\begin{eqnarray}
u_k & = & \sum_{j=1}^m \omega_{kj}x_j
\end{eqnarray} where $\omega_{kj}x_j$ for each $j$ consists of the multiplication of the synaptic weight $\omega$ and a signal $x$, that defines its strength. Thus, an output $y_k$ is produced by means of an activation function $\phi$:
\begin{eqnarray}
y_k & = & \phi(u_k + b_k)
\end{eqnarray} where $b_k$ indicates the bias, which has the effect of increasing or lowering the net input of the activation function. As we aim to make our network accountable for non-linear dependencies, the activation functions need to be also non-linear, such as the log sigmoid or the hyperbolic tangent sigmoid functions. Nevertheless, this choice is user-defined and may depended on the application. A scheme of a neuron within an ANN is given in Fig. \ref{fig1}.

The definition of the weights and biases in the network is defined in the learning phase of the algorithm, and they are adjusted iteratively based on the data given as input-output that is seen by the network. This process aims to minimize the loss (or performance) function which can be, for instance, the squared error between the output of the network and the real output value, and aims to make the network perform in an expected way. The widely used framework in the learning algorithm of an ANN is the gradient descent backpropagation, that update weights and biases in the direction of the negative gradient of the loss function. 

Several ANNs architectures that are based on layers of the neurons are possible. A feedforward network consists of a connection between input layer (where the data is gathered), hidden layers and the output layer (which gives us our desired result). The term 'hidden' refers to the fact that this part of the network is not seen directly either by the input and the output layers \cite{haykin2009neural}. We can consider a single-layer architecture, where no hidden layers are used. The benefit of hidden layers is to give the network a global perspective due to the extra set of synaptic connections and dimension of neural interaction \cite{churchland1994computational}. 

The main characteristic of feedforward networks is that the learning process flows from the input to the output layers, only in this direction, without going back the layers. If the network has at least one feedback loop, forming a cycled connection between layers, then we create recurrent neural networks (RNNs), an architecture of ANNs that aims to give a better accuracy on predictions that needs a memory along the network.

While feedforward ANNs can only map from input to output, RNNs can consider the entire history between inputs and the desired outputs. This means that a neuron's output can be feedback as an input to all neurons of the net. A self-feedback network occurs when an output neuron feeds its own input, while a no self-back network means otherwise, that is, the output of one neuron is used as input to all other neurons but itself. This attribute of RNNs allows a memory of previous inputs to persist in the network’s internal state, and thereby influence the network output \cite{graves2008supervised}. RNNs can be derived from nonlinear first-order non-homogeneous ordinary differential equations; a deep and elucidate analysis can be found in Sherstinsky \cite{sherstinsky2020fundamentals}.

Nevertheless, the use of RNNs in its standard configuration to account for contextual information is still limited, due to the effect known as vanishing gradient problem \cite{hochreiter2001gradient}. As the information circles around the recurrent network in time, the influence of a input on the hidden layer, and consequently on the output, either decays or blows up exponentially.  One attempt to solve this problem is with the Long Short-Term Memory (LSTM) architecture, presented for the first time by Hochreiter and Schmidhuber \cite{hochreiter1997long}.

With the purpose of solving the vanishing gradient problem, long short-term memory (LSTM) architecture incorporates non-linear, data-dependent controls into the RNN cell, so that the gradient of the loss function does not vanish \cite{sherstinsky2020fundamentals}. As we saw in the previous section, RNNs architecture are formed within a circle context, with repeating the modules of the net. The difference for LSTMs is that, although the same strategy of circling information is maintained, the summation in the hidden layer is replaced by a memory block, which has four neural network connected and interacting together. This structure allows LSTMs to learn and remember information for a long-time period, which is its default behaviour.

The architecture of LSTMs is build as follow: inside the memory block, there is one or more central cells that are self-looped into three multiplicative units called input, output and the forget gates. This difference of having more units controls the flow of information \cite{goodfellow2016deep}, where the multiplicative input protects the memory block from receiving perturbation from irrelevant inputs, while the output gates protect other units from irrelevant information of the current block \cite{hochreiter1997long}. The units work as gates to avoid weight conflicts, i.e., the input gate decides when to keep or exclude information within the block, while the output gate decides when to access the block and prevent other blocks from being perturbed by itself.

Figure~\ref{fig2} shows an overview of the memory block inside a LSTM network. The three gates receive activations from both inside and outside (other memories blocks), controlling the activation of that cell by multiplications. The input and output gates multiply the input and output of the cell while the forget gate multiplies the previous state of the cell. If $n$, $m$ and $k$ correspond to the number of inputs, outputs and cells in the hidden layer, respectively, the activation of the input gate $b_{\sigma}^t$ at time $t$ is given by

\begin{eqnarray}
b_{\sigma}^t & = & \phi \left(  \sum_{i = 1}^n \omega_{i \sigma}x_i^t + \sum_{h = 1}^m \omega_{h \sigma}b_h^{t-1} + \sum_{c = 1}^k \omega_{c \sigma}s_c^{t-1}  \right),
\end{eqnarray} where $\sigma$ represents the input gate, $x$ the signal, $\omega$ the weights that will connect two units and $s_c^t$ is the activation of the cell $c$ at time $t$ unit. Usually, the gate activation function $\phi$ is the logistic sigmoid so that the values are between $0$ and $1$ \cite{hochreiter1997long, graves2008supervised, goodfellow2016deep}. The activation of the forget gate $b_\tau^t$ at time $t$ is given by

\begin{eqnarray}
b_\tau^t & = & \phi \left(  \sum_{i = 1}^n \omega_{i \tau}x_i^t + \sum_{h = 1}^m \omega_{h \tau}b_h^{t-1} + \sum_{c = 1}^k \omega_{c \tau}s_c^{t-1}  \right).
\end{eqnarray} while the activation of the output gate $b_\gamma^t$ at time $t$ is

\begin{eqnarray}
b_\gamma^t & = & \phi \left(  \sum_{i = 1}^n \omega_{i \gamma}x_i^t + \sum_{h = 1}^m \omega_{h \gamma}b_h^{t-1} + \sum_{c = 1}^k \omega_{c \gamma}s_c^{t}  \right).
\end{eqnarray}

The activation of the cell $s_c^t$ at time $t$ is

\begin{eqnarray}
s_c^t & = & b_\tau^t s_c^{t-1} + b_{\sigma}^t g \left(   \sum_{i = 1}^n \omega_{i c}x_i^t + \sum_{h = 1}^m \omega_{h c}b_h^{t-1}  \right)
\end{eqnarray} where $g$, the activation function, is usually a hyperbolic tangent of logistic sigmoid functions. Finally, the cell output $b_c^t$ at time $t$ is given by

\begin{eqnarray}
b_c^t & = & b_\gamma^t h(s_c^t)
\end{eqnarray} where $h$ can be the same function as $g$, or even the identity function. In the formulae above, $\omega_{c \sigma}$, $\omega_{c \tau}$ and $\omega_{c \gamma}$ indicate the weights from the cell to the input, forget and output gates, respectively. Note that if the input gate has an activation near zero, it will not open, and therefore the activation of the cell will not be overwritten by new inputs, and will be available later in the sequence, i.e., the block store information for a longer time than usual \cite{graves2008supervised}. 

The calculations explained above are provided for the forward pass of a LSTM hidden layer, starting at $t = 1$ and applying the equations iteratively to update $t$, until the lenght of the input sequence of data is achieved. However, to complete the calculations, a backpropagation backward pass is necessary, starting at the length of the input and going back until $t = 1$. The formulae of this phase can be found in \cite{hochreiter1997long, graves2008supervised, goodfellow2016deep}.

\begin{figure}[h]
\begin{center}
\begin{tikzpicture}[roundnode/.style={circle, draw=black!60, fill=white!5, very thick, minimum size=12mm}, roundsmall/.style={circle, draw=black!60, fill=white!5, thick, minimum size=4mm}, thickround/.style={circle, draw=black!60, fill=white!5, ultra thick, minimum size=15mm}, squarednode/.style={rectangle, draw=black, fill=black, very thick, minimum size=5mm}, box/.style = {draw, rounded corners=15pt, minimum width=100mm, minimum height=90mm}], node distance = 2cm]
    
    \node (a)   [box] {};
    \node[thickround]       (cell)      {\textbf{CELL}};
    \node[roundsmall]       (acth)      [below = of cell] {$h$};
    \node[squarednode]      (black1)    [below left = of cell] {};
    \node[roundnode]        (circle1)   [label=below: Input gate] [left = of black1] {$b_{\sigma}^t$};
    \node[squarednode]      (black2)    [right = of acth] {};
    \node[roundnode]        (circle2)   [label=below: Output gate] [right = of black2] {$b_{\gamma}^t$};
    \node[squarednode]      (black3)    [above = of cell] {};
    \node[roundnode]        (circle3)   [label=left: Forget gate] [left = of black3] {$b_{\tau}^t$};
    \node[roundsmall]       (actg)      [below = of black1] {$g$};
    \node[]                 (out1)      [below = of actg] {};
    \node[]                 (kjk)       [above = of circle3] {};
    \node[]                 (out2)      [above = of kjk] {};
    \node[]                 (lala2)     [below = of black2] {};
    \node[]                 (out3)      [below = of lala2] {};
    \node[]                 (out4)      [left = of circle1] {};
    \node[]                 (out5)      [right = of circle2] {};

    \draw[->, shorten >= 5pt, shorten <= 5pt, line width=0.4mm] (circle1) edge (black1);
    \draw[->, shorten >= 5pt, shorten <= 5pt, line width=0.4mm] (circle2) edge (black2);
    \draw[->, shorten >= 5pt, shorten <= 5pt, line width=0.4mm] (circle3) edge (black3);
    \draw[->, shorten >= 5pt, shorten <= 5pt, line width=0.4mm] (cell) edge[bend left] (black3);
    \draw[->, shorten >= 5pt, shorten <= 5pt, line width=0.4mm] (black3) edge[bend left] (cell);
    \draw[dashed, ->, shorten >= 5pt, shorten <= 5pt, line width=0.4mm] (cell)  edge node [left] {$\omega_{c \tau}$} (circle3);
    \draw[dashed, ->, shorten >= 5pt, shorten <= 5pt, line width=0.4mm] (cell)  edge[bend right] node [left] {$\omega_{c \sigma}$} (circle1.north);
    \draw[dashed, ->, shorten >= 5pt, shorten <= 5pt, line width=0.4mm] (cell)  edge[bend left] node [right] {$\omega_{c \gamma}$} (circle2.north);
    \draw[->, shorten >= 5pt, shorten <= 5pt, line width=0.4mm] (black1) edge (cell);
    \draw[->, shorten >= 5pt, shorten <= 5pt, line width=0.4mm] (acth) edge (black2);
    \draw[->, shorten >= 5pt, shorten <= 5pt, line width=0.4mm] (cell) edge (acth);
    \draw[->, shorten >= 5pt, line width=0.4mm] (out1) edge (actg);
    \draw[->, shorten >= 5pt, shorten <= 5pt, line width=0.4mm] (actg) edge (black1);
    \draw[->, shorten >= 5pt, shorten <= 5pt, line width=0.4mm] (out2) edge (circle3);
    \draw[->, shorten >= -5pt, shorten <= 5pt, line width=0.4mm] (black2) edge (out3);
    \draw[->, shorten >= 5pt, shorten <= -5pt, line width=0.4mm] (out4) edge (circle1);
    \draw[->, shorten >= 5pt, shorten <= -5pt, line width=0.4mm] (out5) edge (circle2);

\end{tikzpicture}
\end{center}
\caption{LSTM memory block. The input, output and forget gates that controls the activation of the cell through the multiplicative units (black squares) from inside and outside. Note that the forget gates will multiply the previous state of the cell. Functions $g$ and $h$ are activation functions explained in the text.}
\label{fig2}
\end{figure}

\section{Data and area of study} \label{data}
The LSTM methodology described in the previous section is used here to predict the significant wave height $H_s$, based on the ERA5 reanalysis and observational data. Values of $H_s$ are gathered from ERA5 hourly from 1979 to present \cite{copernicus, hersbach2020era5}. ERA5 is the fifth generation ECMWF reanalysis for the global climate and weather for the past 4 to 7 decades. In our experiments, the training dataset is selected until one month prior to the predictions period (see Tab. \ref{tab:1}).

\begin{figure}
    \centering
    \includegraphics[scale=0.8]{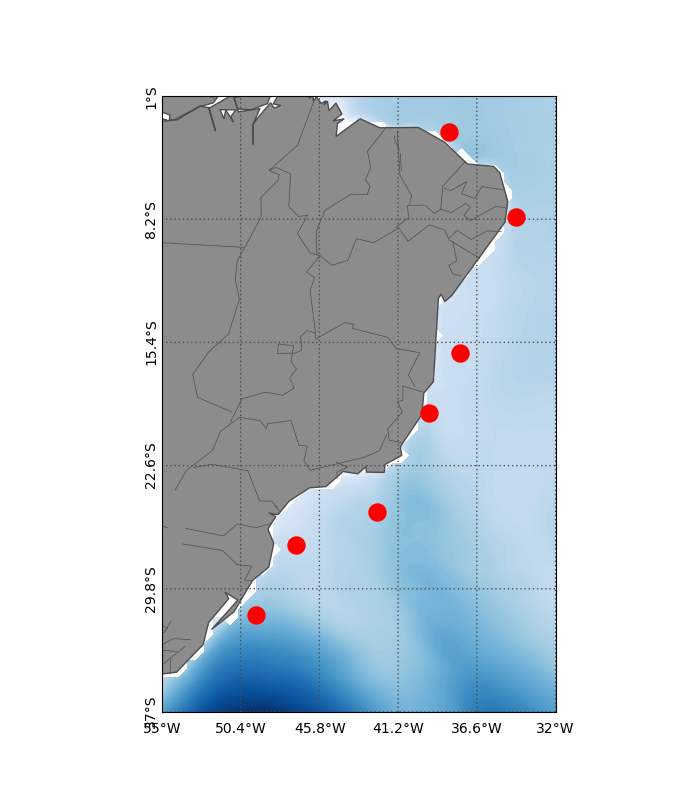}
    \caption{Region of analysis in the Brazilian coast. Red circles indicates the location of the seven buoys studied in this work.}
    \label{fig:3}
\end{figure}

Seven buoys locations are considered for this study. All of them are located near in the Brazilian coast, ranging from longitude 49° 86'W to 38° 25'W and latitudes 31° 33'S to 3° 12'S. These buoys belong to the National Program of Buoys (PNBOIA) of the Brazilian Navy, which aims to collect oceanographic and meteorological data of the Atlantic Ocean \cite{pereira2017wave, pnboia}. Figure~\ref{fig:3} shows the region of analysis, with red circles indicating the location of the buoys, while Tab. \ref{tab:1} presents longitudes and latitudes of the seven buoys. Predictions are made for verification times within one month, but the month chosen vary for each buoy, due to the lack of complete data in the databases. Table~\ref{tab:1} also shows the period of predictions to each buoy. Nowadays, only buoy 7 is still operating, while all others are in maintenance.

In this work, we use three metrics to analyse the accuracy of our results. The well known mean absolute error (MAE), which is given by
\begin{eqnarray}
MAE & = & \frac{1}{n} \sum_{i = 1}^n \left| \tilde{y}_i - y_i \right|,
\end{eqnarray} where the tilde means true value while non-tilde means predicted value ($n$ is the number of observations). The relative error (RE) for each step is given by
\begin{eqnarray}
RE & = & 100\frac{\left| \tilde{y}_i - y_i \right|}{\left| \tilde{y}_i \right|},
\end{eqnarray} and the mean absolute relative error (MAPE),
\begin{eqnarray}
MAPE & = & 100 \frac{1}{n} \sum_{i = 1}^n \frac{\left| \tilde{y}_i - y_i \right|}{\left| \tilde{y}_i \right|}.
\end{eqnarray} Both MAPE and RE are given in percentages while MAE is in the same unit as the data.

\begin{table}
\scriptsize
    \begin{tabular}{l|c|c|c|c|c}
         & Longitude & Latitude & Period of prediction & Buoy's depth & City/State location \\ \hline
        Buoy 1 & 49° 86' W & 31° 33' S  & March/2018 &  200m & Rio Grande/RS \\
        \hline 
        Buoy 2 & 47° 15' W & 27° 24' S  & March/2018 & 200m  & Itajaí/SC \\
        \hline 
        Buoy 3 & 42° 44' W & 25° 30' S  & January/2021 & 2164m & Santos/SP \\
        \hline 
        Buoy 4 & 39° 41' W & 19° 55' S  & March/2017 & 200m & Vitória/ES \\
        \hline 
        Buoy 5 & 37° 56' W & 16° 00' S  & March/2016 & 200m & Porto Seguro/BA \\
        \hline 
        Buoy 6 & 34° 33' W & 8° 09' S & March/2016 & 200m & Recife/PE \\
        \hline 
        Buoy 7 & 38° 25' W & 3° 12' S & March/2017 & 200m  & Fortaleza/RN \\
    \end{tabular}
    \caption{Geo-spatial latitude and longitude location of the seven buoys used in this work, period of prediction, buoy's depth and city of location in Brazil.}
    \label{tab:1}
\end{table}

\begin{figure}
    \centering
    \includegraphics[scale=0.6]{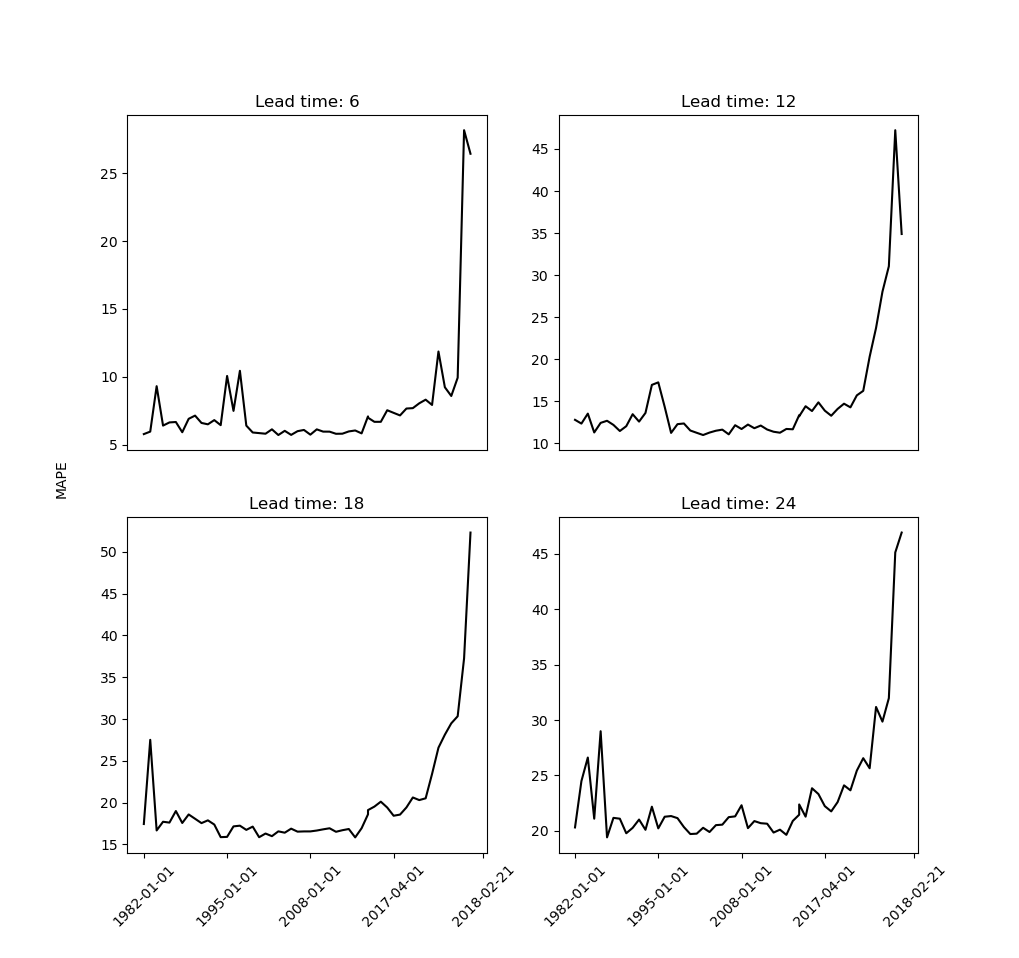}
    \caption{MAPE calculated for LSTM predictions using different starting dates: from January/1979 until February/2018. Simulations were performed for buoy number 1.}
    \label{fig:4}
\end{figure}

Since the training dataset time span can have impact in the predictions, both on accuracy and processing time, we conducted an experiment with the data gathered from 1979 until 2018, to find that, for different lead times, the amount of data needed for a better accuracy is different. We show in Fig. \ref{fig:4} the MAPE for starting the date of training set from 01-01-1979 to 28-03-2018 (see the $x-$ axis), i.e, data range from 39 years to one week prior to the prediction period. Simulations and the establishment of the training dataset to be used adopted exclusively buoy number one (Rio Grande). As we can see, the MAPE starts to grow exponentially with a training dataset of only one year, from February/2017 until the end of February/2018, which is an expected behaviour, considering the need for historical data of the algorithm. However, as the training set becomes larger than one year, we see that although LSTM is developed to remember historical data, is not that the bigger the dataset, the better is the accuracy. In fact, for all lead times, there is an increase of MAPE in some datasets that are longer than 25 years. This could happen because of the quality of the data or how the variable is changing over time, and is object for further investigation.


Thus, to reach the better results for a prediction, we consider the accuracy, size of the training set and global accuracy, both by MAPE, MAE and RE. Based on this analysis, we set the starting date of the training sets for each lead time differently, as summarized in Table~\ref{tab:2}. Still, it is possible to reach good accuracy with one year of dataset; simulations will run fast and the predictions will be satisfactory, although with some loss of accuracy if compared with the training size datasets we have chosen in the work. Its important to mention also that, with smaller datasets, the larger lead time predictions, specially eighteen and twenty four, will fail to predict peaks or valleys since the network will approach the average value of the period.

\begin{table}
    \centering
    \begin{tabular}{l|c|c}
         Lead & MAPE & Training start date  \\
         \hline
         6 & $5.70\%$ & 01-01-2003 \\ \hline
         12 & $10.98\%$ & 01-01-2002 \\ \hline
         18 & $15.85\%$ & 01-01-2005 \\ \hline
         24 & $19.41\%$ & 01-01-1987 \\ 
    \end{tabular}
    \caption{Training set starting date best MAPE accuracy results for four different lead times considered in this work.}
    \label{tab:2}
\end{table}
\section{Methodology} \label{method}
\begin{figure}[h]
\begin{center}
\begin{tikzpicture}[
roundsmall/.style={circle, draw=black, fill=black, thick, minimum size=1mm, inner sep=2pt},
squarednode/.style={rectangle, draw=black, fill=black!30, very thin, minimum width=30mm, minimum height=9mm}, 
squarednode2/.style={rectangle, draw=black, fill=white, very thin, minimum width=20mm, minimum height=9mm}, 
squarednode3/.style={rectangle, draw=black, fill=black!30, very thin, minimum width=42.5mm, minimum height=9mm}, 
squarednode4/.style={rectangle, draw=black, fill=black!30, very thin, minimum width=52.5mm, minimum height=9mm},
squarednode5/.style={rectangle, draw=black, fill=black!30, very thin, minimum width=62.5mm, minimum height=9mm}, 
squarednode6/.style={rectangle, draw=black, fill=black!30, very thin, minimum width=90mm, minimum height=9mm}],node distance = 0.1cm]
    
    \node (a)   [squarednode] {Training dataset};
    \node (b)   [squarednode2] [right = 0.1cm of a] {\footnotesize Lead time 6};
    \node (c)   [roundsmall] [right = 0.1cm of b] {};
    \node (d)   [roundsmall] [right = 5.5cm of c] {};
    \node (e)   [] [above = 0.2cm of c] {$t_0$};
    \node (f)   [] [above = 0.2cm of d] {$t_n$};
    
    \node (i)   [roundsmall] [below right = 1.2cm of c] {};
    \node (g)   [squarednode2] [left = 0.1cm of i] {\footnotesize Lead time 6};
    \node (h)   [squarednode3] [left = 0.1cm of g] {Training dataset};
    \node (j)   [roundsmall] [below = of d.east, anchor=east] {};
    \node (k)   [] [above = 0.2cm of i] {$t_1$};
    \node (l)   [] [above = 0.2cm of j] {$t_n$};
    
    \node (m)   [roundsmall] [below right = 1.2cm of i] {};
    \node (n)   [squarednode2] [left = 0.1cm of m] {\footnotesize Lead time 6};
    \node (o)   [squarednode4] [left = 0.1cm of n] {Training dataset};
    \node (p)   [roundsmall] [below = of j.east, anchor=east] {};
    \node (q)   [] [above = 0.2cm of m] {$t_2$};
    \node (r)   [] [above = 0.2cm of p] {$t_n$};
    
    \node (s)   [roundsmall] [below right = 1.2cm of m] {};
    \node (t)   [squarednode2] [left = 0.1cm of s] {\footnotesize Lead time 6};
    \node (u)   [squarednode5] [left = 0.1cm of t] {Training dataset};
    \node (v)   [roundsmall] [below = of p.east, anchor=east] {};
    \node (x)   [] [above = 0.2cm of s] {$t_3$};
    \node (z)   [] [above = 0.2cm of v] {$t_n$};
    
    \node (aa)   [] [below = 0.1cm of u] {$\vdots$};
    \node (ab)   [] [below = 0.1cm of t] {$\vdots$};
    \node (ac)   [] [below = 0.45cm of s] {$\vdots$};
    \node (ad)   [] [below = 0.45cm of v] {$\vdots$};
    
    \node (ae)   [roundsmall] [below = 1.5cm of ad.east, anchor=east] {};
    \node (af)   [squarednode2] [left = 0.1cm of ae] {\footnotesize Lead time 6};
    \node (ag)   [squarednode6] [left = 0.1cm of af] {Training dataset};
    \node (aj)   [] [above = 0.2cm of ae] {$t_n$};
    
    \draw [decorate, decoration={brace,amplitude=7pt}] (e.north) -- (f.north) node [midway,yshift=0.3in] {prediction period};
    \draw[dashed, line width=0.4mm] (c) -- (d);
    \draw[dashed, line width=0.4mm] (i) -- (j);
    \draw[dashed, line width=0.4mm] (m) -- (p);
    \draw[dashed, line width=0.4mm] (s) -- (v);

\end{tikzpicture}
\end{center}
\caption{Schematic of the methodology for training applied in this work, for lead time six (other lead times follow analogously). For each verification time $t_i$, where $i = 1, \ldots, n$, the training set consists of data until six hours steps prior to $t_i$.}
\label{fig:lead}
\end{figure}
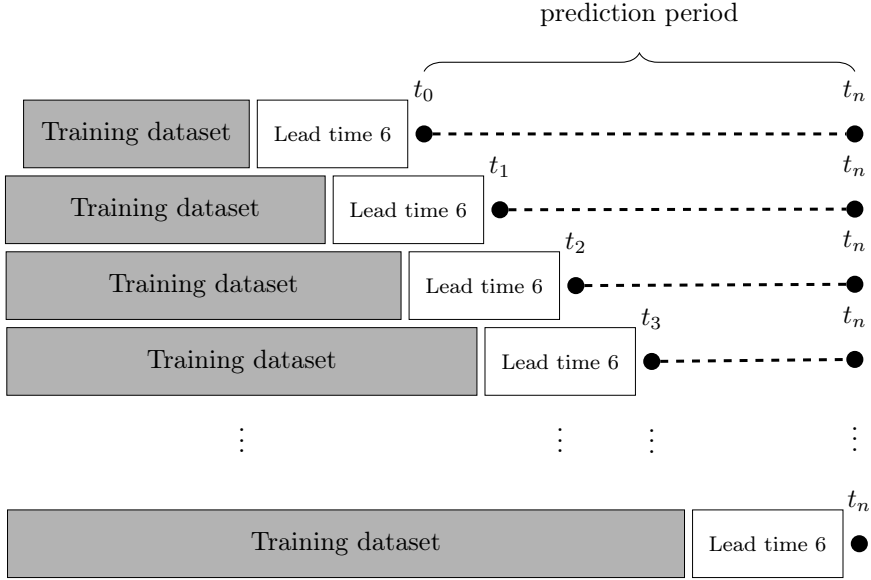

The main goal of this work is to predict the significant wave height $H_s$ for the seven buoys locations presented in Table~\ref{tab:1}, hourly for four lead times (6, 12, 18 and 24), using the LSTM neural network. As mentioned, the starting date of training set is different for each lead time, and is defined as showed in Table~\ref{tab:2}, while the end of the database depends on the lead time, that is, for each verification time, a new training is performed with the training set consisting of data from the start date until the prediction time minus the lead time. With this, we guarantee that our training will have as much as data necessary for the best result, respecting the lead time gap. Figure~\ref{fig:lead} shows a schematic of the training methodology applied.

We considered different strategies for the prediction: first, only the historical series of $H_s$ in the buoy location is used for training. We also investigate the influence of other variables in the training set, based both on physical aspects and on a correlation analysis. Thus, predictions of $H_s$ is also performed using  other physical variables, but excluding $H_s$, as input. Lastly, we do not use ERA5 reanalysis data as input, but the real buoy data observations to train the network. These different frameworks are explained in more detail in section~\ref{results}.

In the training phase of the LSTM, a cross-validation scheme was implemented, where $80\%$ of the data is selected for the training and $20\%$ for validation. This strategy is an excellent framework to avoid overfitting of a model, i.e., a model that yields a good accuracy to the validation set (seen data) and a bad result to unseen data. Thus, in our methodology, the data was divided into the effective training set, the validation set ($80/20$ split) and the set of values to predict, called test set, using a standard terminology from machine learning  algorithms. 

To avoid excess in the memory usage of the training phase, the full set of data is subdivided into smaller batches. Training is then performed for each batch, with the target value considered for this batch, and respecting the lead time. To define the batch size, several simulations were performed, and a optimal value of twelve data points was obtained.

The Python library TensorFlow \cite{tensorflow2015-whitepaper} is an end-to-end open source platform for machine learning and its Keras API \cite{chollet2015keras} are used in this work to implement the LSTM algorithm. The model is compiled using the mean absolute error as loss function which is optimized by the Adam algorithm. Adam optimization is a stochastic gradient descent method that is based on adaptive estimation of first-order and second-order moments. The LSTM network is build with three layers, one with 64, other with 48 neurons and the last with 32. These values yields the best results possible, considering possible hyper-parameters to tune the model, considering the metric (MAPE) and the computational time.
\section{Results} \label{results}
In this section, we present the results of the predictions for each buoy location and lead time. Simulations are performed using both GPU and CPU parallelization, to improve performance. Early stopping is used for convergence of the solution, i.e., as the learning rate achieves a desired accuracy based on MAE, the training stops.

For each lead time, our algorithm took approximately 14 minutes for training the LSTM and the prediction of a single time value took $2.62 \times 10^{-6}$ seconds. The simulations were performed in a machine with Intel Xeon processor with $20$ cores, $128$ Gb of RAM memory, with a GeForce RTX 2080 Ti graphic card. The GPU is used to parallelize  the training process.
\subsection{Predictions using $H_s$ both as feature and target}

\begin{figure}
    \centering
    \includegraphics[scale=0.7]{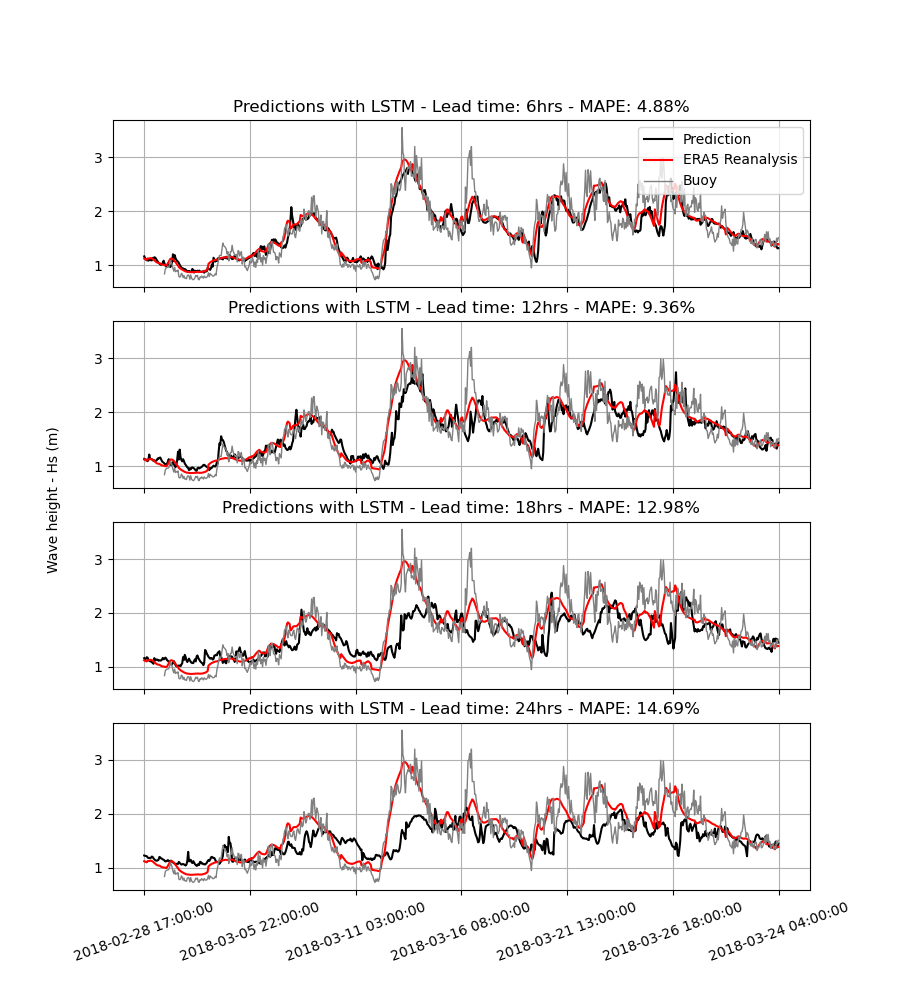}
    \caption{Predictions (black) and reanalysis from ERA5 (red) comparison for buoy location number two. The blue dots show real observations obtained from the buoy.}
    \label{fig:5}
\end{figure}

\begin{figure}
    \centering
    \includegraphics[scale=0.7]{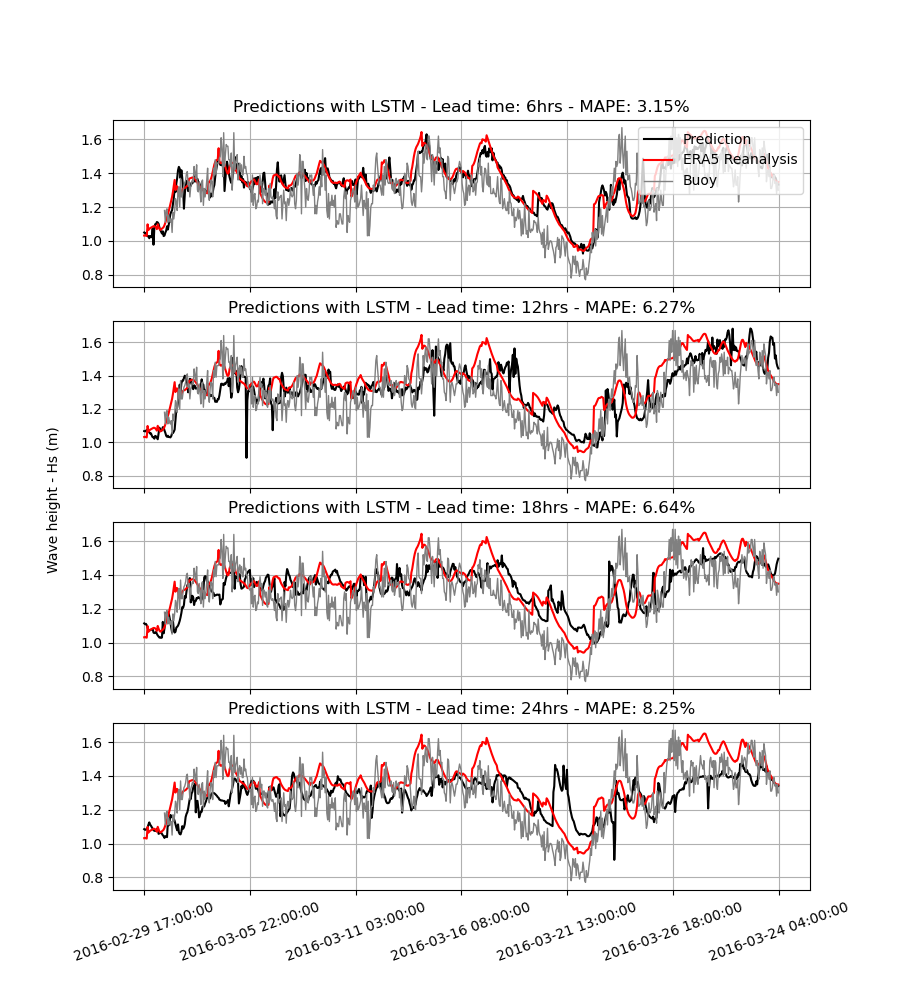}
    \caption{Predictions (black) and reanalysis from ERA5 (red) comparison for buoy location number five. The grey dashed line show real observations obtained from the buoy.}
    \label{fig:7}
\end{figure}

\begin{figure}
    \centering
    \includegraphics[scale=0.7]{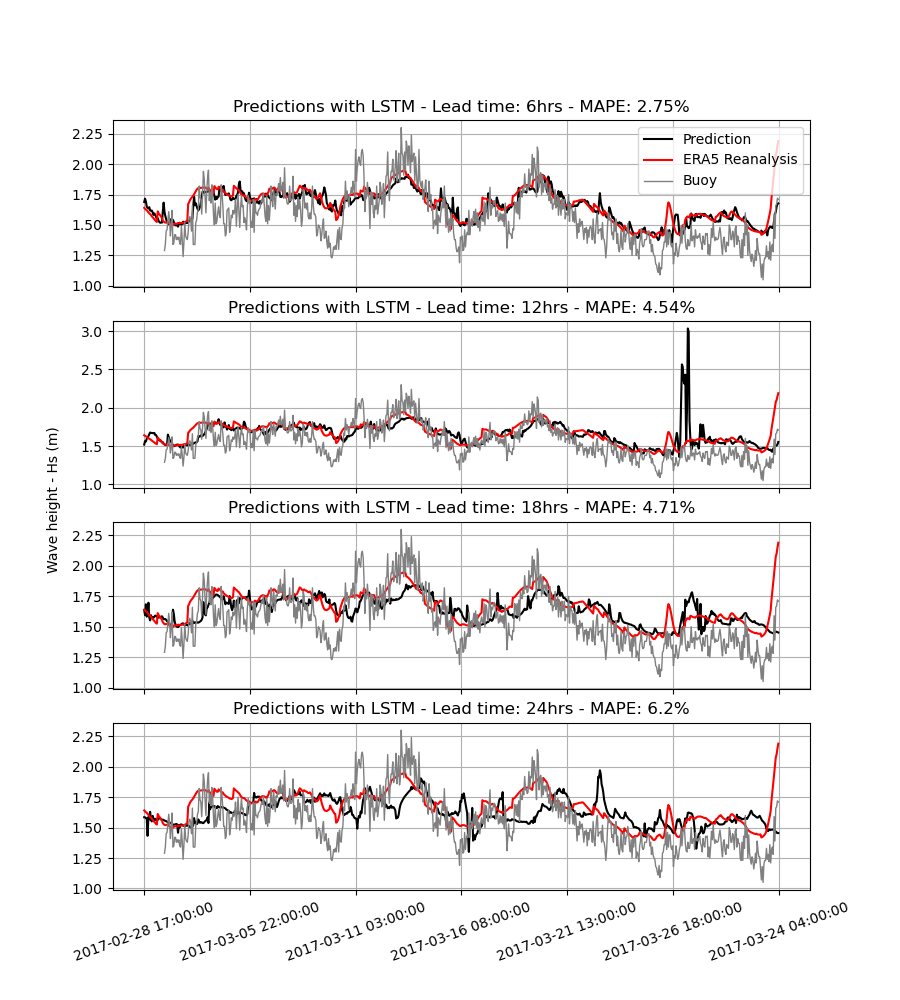}
    \caption{Predictions (black) and reanalysis from ERA5 (red) comparison for buoy location number seven. The grey dashed line show real observations obtained from the buoy.}
    \label{fig:9}
\end{figure}

\begin{figure}
    \centering
    \includegraphics[scale=0.55]{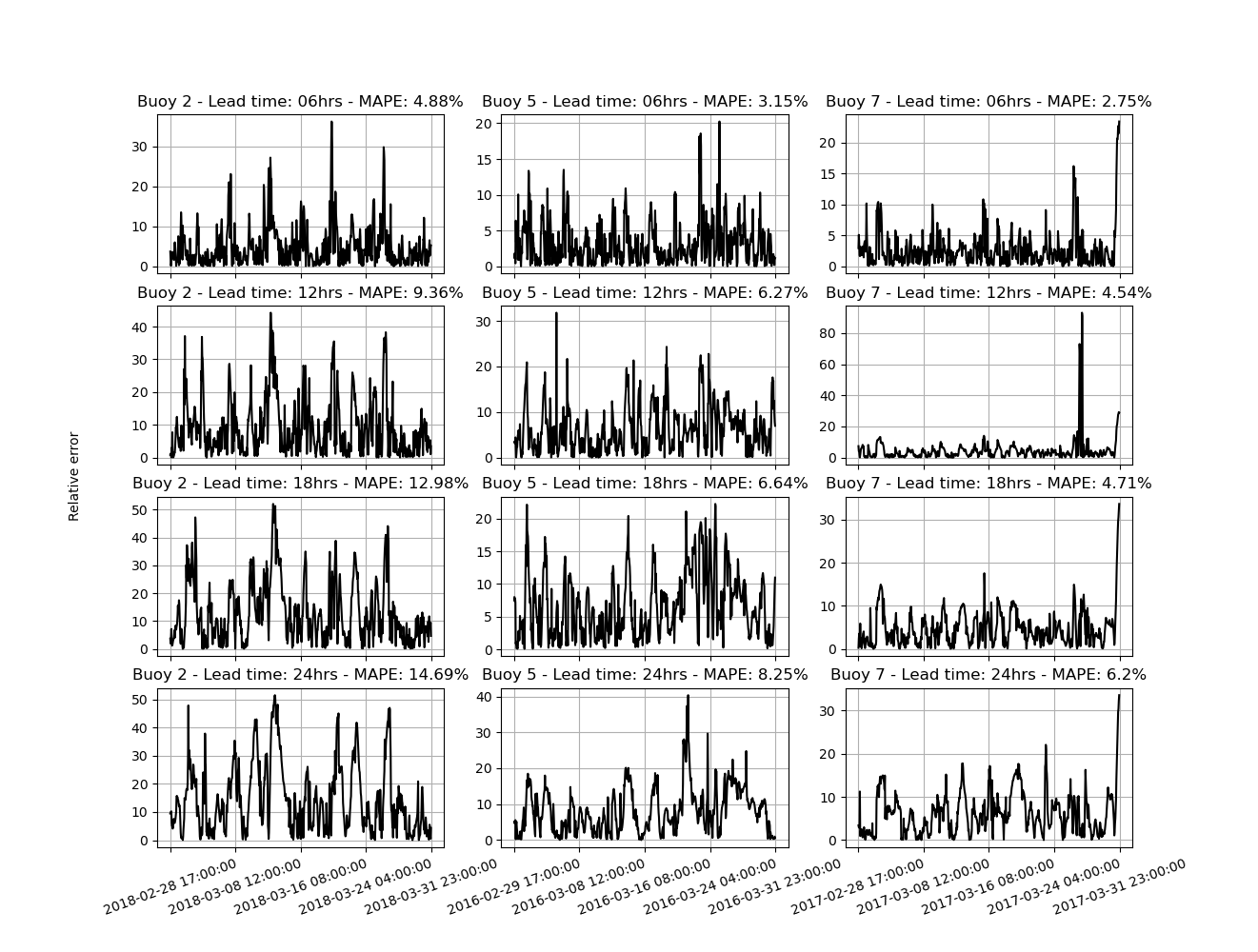}
    \caption{Relative error between predicted value and ERA5 reanalysis for buoys locations number two, five and seven.}
    \label{fig:6}
\end{figure}

Figures~\ref{fig:5}, \ref{fig:7} and~\ref{fig:9} present the comparison between predictions obtained with LSTM, the ERA5 reanalysis data (same data in the four figures) and the observations, for buoys locations number two, five and seven, respectively. As it can be observed, lead time 6 yields a lower MAPE,  calculated between LSTM prediction and ERA5 reanalysis data, in all the locations, which is expected since the gap between the training and the verification times is smaller, so the memory that the network needs to keep is smaller. For buoy location number two, based on the MAPE, the results show an accuracy\footnote{Here, the accuracy is obtained subtracting MAPE from $100\%$} of $95.12\%$ for lead time 6, while accuracy is $90.64\%$ for lead time 12, accuracy of $87.02\%$ for lead time 18 and accuracy of $85.31\%$ for lead time 24. We can see that the pattern for lead time 6 is very well described, with the forecast series closely following the reanalysis values and the buoy data. Except for two times. e.g., 2018-03-20 at 1:00:00 UTC and 2018-03-20 at 2:00:00 UTC, where the relative error (see Fig. \ref{fig:6}) is above $30\%$, all errors are below $25\%$ for lead time $6$ whilst the MAE is equal to a remarkable $0.086$ m. Despite a small loss of accuracy for lead times 12, 18 and 24, we still observe an accuracy of more than $85\%$ for all lead times in this location.

Predictions for buoys locations number five and seven show the best accuracies in all studied buoys locations, as we can see in Table~\ref{tab:3}. An outstanding accuracy of $97.25\%$ is obtained for lead time 6 in buoy location number seven, while the 24h lead time prediction, expected to behave somewhat poorly, has an accuracy of $93.8\%$. The predicted values follows closely the reanalysis data, and the reason for this pattern could be explained by the fact that the historical series of reanalysis has a high consistency in its values, considering the seasonality of the series that facilitates the training process, leading to a better optimization of the error in the network. The results for buoy location five also have an accuracy above $90\%$ for all lead times, and the same quality of the historical series is observed. One can see the relative error in Figure~\ref{fig:6} for buoys locations numbers two, five and seven, which corroborates with the MAPE metrics.

\begin{table}
\small
    \centering
    \begin{tabular}{l|c|c|c|c|c|c|c|c|}
    & \multicolumn{4}{|c|}{MAPE} & \multicolumn{4}{|c|}{MAE} \\ \hline
     & 6 & 12 & 18 & 24 & 6 & 12 & 18 &  24 \\ \hline
        Buoy 1 & $6.12\%$ & $11.9\%$ & $18.82\%$ & $24.15\%$ & 0.13m & 0.26m & 0.39m &  0.50m \\ \hline
        Buoy 2 & $4.88\%$ & $9.36\%$ & $12.98\%$ & $14.69\%$ & 0.08m & 0.16m & 0.23m &  0.26m \\ \hline
        Buoy 3 & $4.28\%$ & $9.06\%$ & $11.83\%$ & $13.26\%$ & 0.08m & 0.17m & 0.23m &  0.25m \\ \hline
        Buoy 4 & $4.67\%$ & $8.87\%$ & $11.34\%$ & $12.27\%$ & 0.06m & 0.12m & 0.15m &  0.17m \\ \hline
        Buoy 5 & $3.15\%$ & $6.27\%$ & $6.64\%$ & $8.25\%$ & 0.04m & 0.08m & 0.08m &  0.11m \\ \hline
        Buoy 6 & $2.19\%$ & $3.7\%$ & $4.97\%$ & $5.98\%$ & 0.03m & 0.05m & 0.07m &  0.09m \\ \hline
        Buoy 7 & $2.75\%$ & $4.54\%$ & $4.71\%$ & $6.2\%$ & 0.04m & 0.07m & 0.08m &  0.10m \\ \hline
         
    \end{tabular}
    \caption{MAPE and MAE metrics for each buoy location and lead times for $H_s$ predictions with LSTM.}
    \label{tab:3}
\end{table}

Some disparity is seen between predictions and observed values by the buoys at locations five and seven. As we can see in Figures~\ref{fig:7} and~\ref{fig:9}, the buoy data have a higher variance from the mean, with values ranging from low heights to higher heights, but not crossing $2$m. This leads the LSTM (and also the reanalysis) to smooth the prediction around a mean value. Nevertheless, the pattern of the observation data is well described by the deep learning algorithm. This behaviour is not seen in buoy location number two, and a possible reason for that being due to the difference in water depths in the regions where the data are obtained. Buoys locations number five and seven are in shallow water while buoy location number two is in deeper water.

\begin{figure}
    \centering
    \includegraphics[scale=0.7]{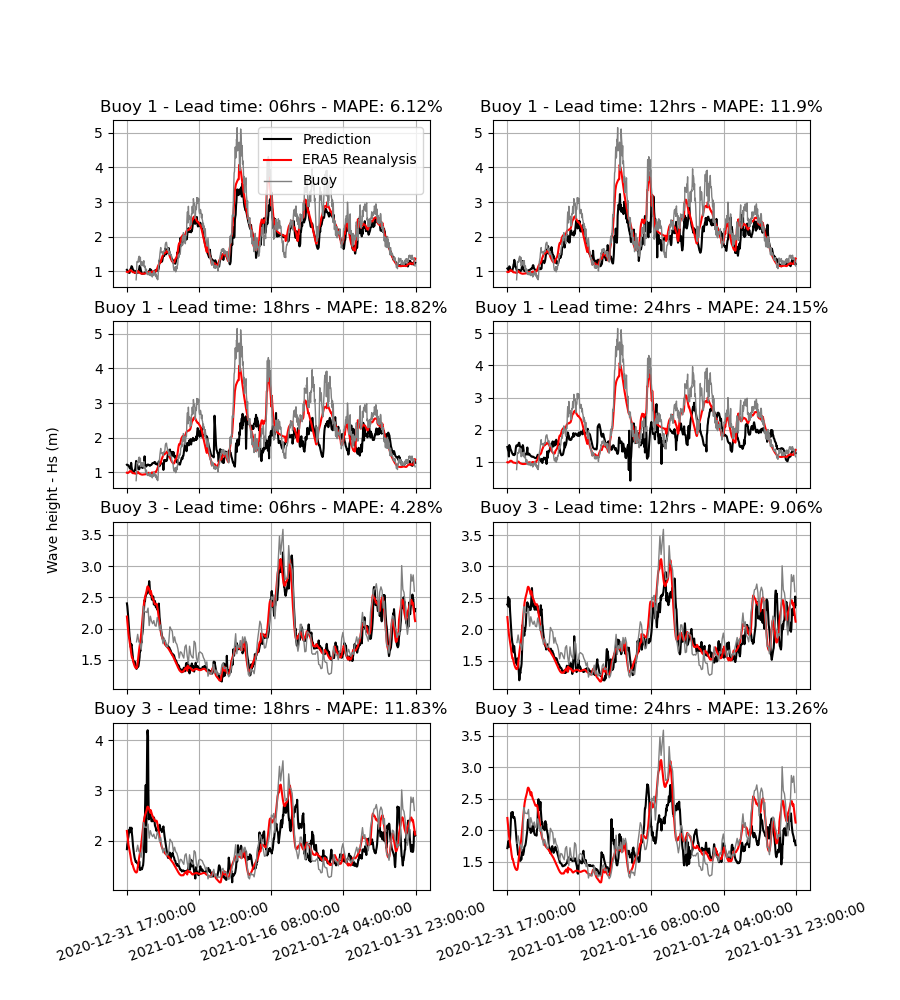}
    \caption{Predictions (black) and reanalysis from ERA5 (red) comparison for buoys locations number one and three. The blue dots show real observations obtained from the buoy.}
    \label{fig:11}
\end{figure}

\begin{figure}
    \centering
    \includegraphics[scale=0.7]{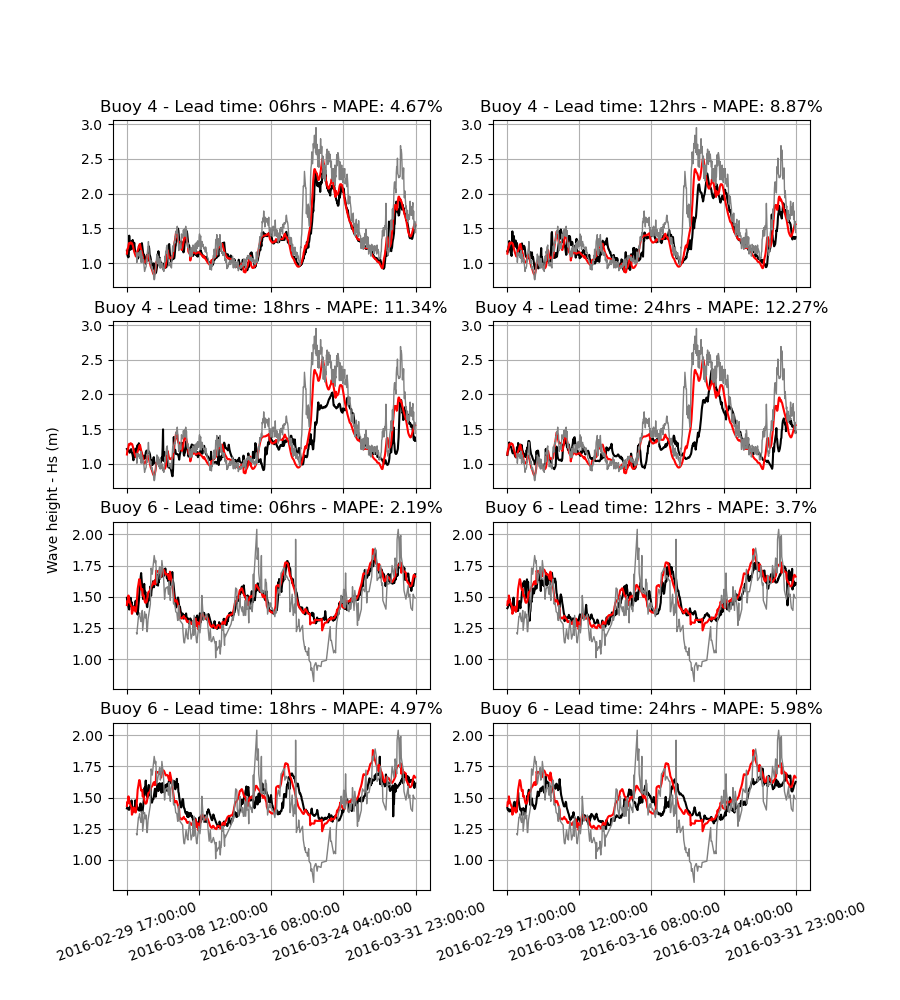}
    \caption{Predictions (black) and reanalysis from ERA5 (red) comparison for buoys locations number four and six. The gray dashed line show real observations obtained from the buoy.}
    \label{fig:12}
\end{figure}

The predicted results for buoys locations number one, three, four and six are shown in Figures~\ref{fig:11} and~\ref{fig:12}. Although the metrics and the accuracy of these results are lower if compared with the three other buoys locations mentioned above, we see a very good behaviour of the LSTM algorithm proposed in this work. Predictions for buoys locations number one and three follow closely the data from reanalysis and observations, while despite the fact that reanalysis and predictions are very similar, a little discrepancy is seen between buoys locations four and six, if compared to observations. This could also be due to the depth of the water in this four buoys locations.

\begin{figure}
    \centering
    \includegraphics[scale=0.7]{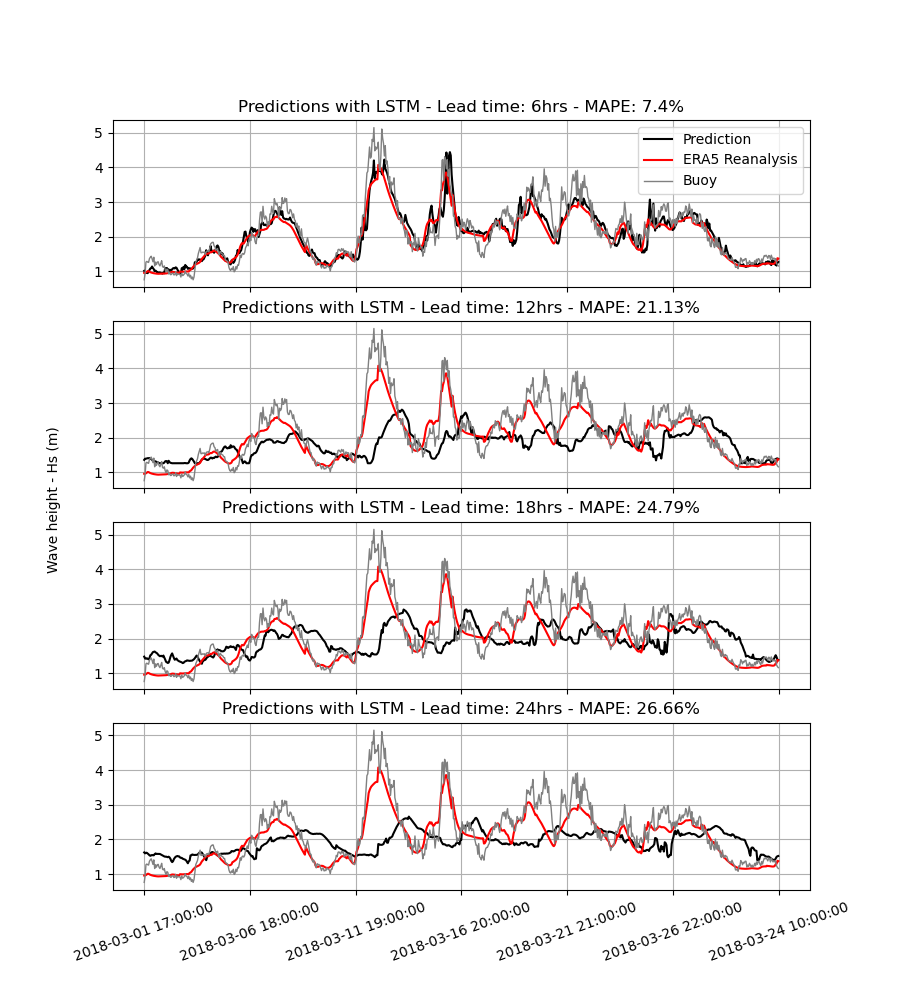}
    \caption{Predictions (black) for buoy location number one and reanalysis from ERA5 (red) comparison for multi variable ($T_p$ and $U_{10}$) simulation. The gray dashed line show real observations obtained from the buoy.}
    \label{fig:13}
\end{figure}

\subsection{Multivariate predictions of $H_s$}
So far we have considered historical series of only significant wave height as input to our model, that is, the LSTM considered only the past values of the variable that the network aims to predict. However, we could consider more than one variable as input to train the network, and also consider one (or multiple) output as the prediction. In this sense, the LSTM will {\em see} how the data from other variables can improve  the predicted variable. We call the variables that will not be predicted, but used as input to the model as features, while the significant height, in this work, will remain our target variable.  In what follows, we decided to conduct our experiments with buoy location number one, since this has a lower global accuracy compared to the others, both for MAPE and MAE. Our multivariate simulation, the architecture of the network and the parameters used are the same as the single variable simulation. The difference here is that two more variables are considered in the input: the peak wave period $T_p$ and the $10$m wind speed $U_{10}$. As output, we have only the prediction of the significant wave height. Note, however, that $H_s$ is also used as input of the model.

We show in Figure~\ref{fig:13} the predicted values of significant wave height for this multivariate version of our algorithm. We can see that there is no considerable improvement in the result and the accuracy of the predictions using two more variables as input. In fact, a large error is observed for lead times $12$, $18$ and $24$, and the MAPE for these lead times are similar, as well as the trend in the graphs. We can conclude here that, the gap between the data used in the training and the effective prediction, i.e, the lead time from $12$, affects the results and creates a pattern that is in fact the same in the predictions. This behaviour, however, is not seen in the lead time $6$, since the gap between the last training data and the prediction is smaller. Thus, based on the accuracy found, we can state that the multi variable simulation has not improved the results from the LSTM network, as already observed in other works in the literature using non-recurrent neural networks \cite{londhe2006one}.

\begin{figure}
    \centering
    \includegraphics[scale=0.52]{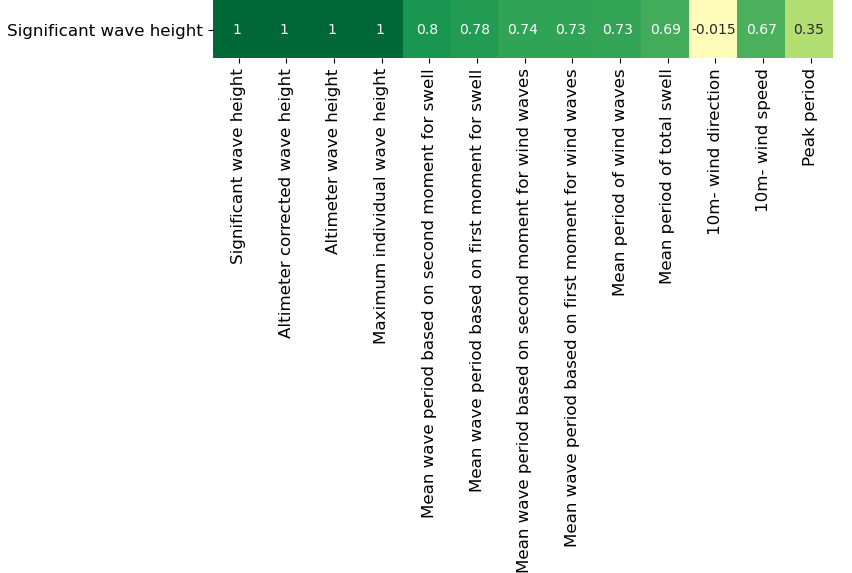}
    \caption{Pearson's coefficient correlation between the best 12 ERA5 ocean wave variables and $H_s$.}
    \label{fig:corr}
\end{figure}

Roughly speaking, in a model based only in data, the best correlated features will, in theory, improve the results in the prediction of the target variable. Features that are physically known to affect the target variable (either linearly or non-linearly) may not improve the accuracy of our model, and can even reduce it comparing to a single variable simulation. This could happen because adding more features as input of your model, the network will try to see the pattern between these values and thus try to correlate then, based on what the historical series show. Therefore, we also developed a multivariate analysis using the best correlated variables from those available in ERA5 database with $H_s$. We use the Pearson's correlation coefficient, where we multiply deviations from the mean for the variable 1 times those for variable 2, and divide by the product of the standard deviations. We have:

\begin{eqnarray}
r_{\text{correlation}} & = & \frac{\displaystyle\sum_{i=1}^n  (x_i - \overline{x})(y_i - \overline{y})}   {(n-1)s_xs_y}.
\end{eqnarray}

The correlation coefficient always lies between one (perfect positive correlation) and minus one (perfect negative correlation); zero indicates no correlation. We show in Figure~\ref{fig:corr} the correlation between twelve variables available in ERA5 and $H_s$. It is important to note that this correlation is based only in data, and does not take into consideration the physical relationship between the variables. As maximum individual, altimeter and altimeter corrected wave height are obviously identically correlated ($r_{\text{correlation}} = 1$) and can bring no improvement to the training, we choose the following four variables for our multivariate simulation:
\begin{enumerate}
    \item Mean wave period based on the second moment for swell;
    \item Mean wave period based on the first moment for swell;
    \item Mean wave period based on the second moment for wind waves;
    \item Mean wave period based on the first moment for wind waves.
\end{enumerate}

\begin{figure}
    \centering
    \includegraphics[scale=0.7]{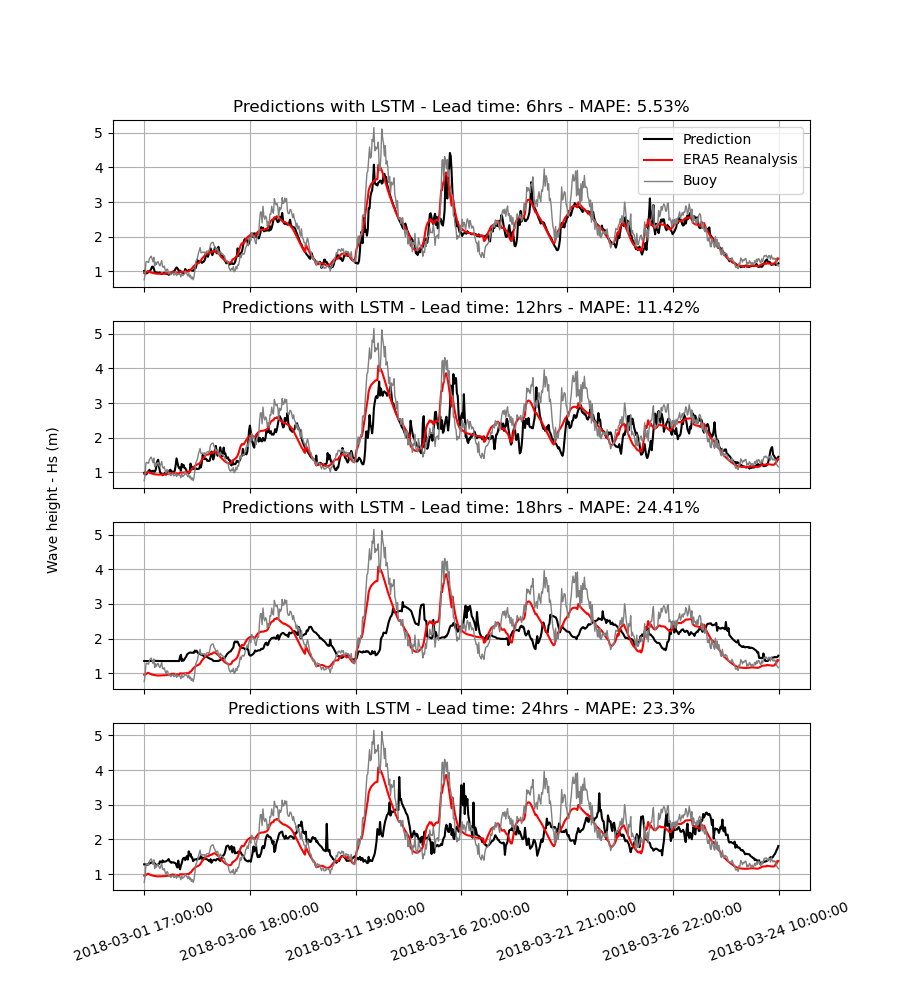}
    \caption{Multi-variable prediction for buoy location number one using LSTM for the best correlated variables. Predictions are in black, reanalysis from ERA5 in red and buoy data given in grey.}
    \label{fig:14}
\end{figure}

We have executed the algorithm with the input training data consisting of five variables: the significant wave height, which is our target variables, and the four enumerated above.  The results are shown in Figure~\ref{fig:14} and as we can see, the they do not improve significantly the previous in this subsection.
\subsection{Predictions not considering $H_s$ as feature}
We have also predicted  $H_s$ using only other physical variables as input (feature), that is, we train our model using $H_s$ as the target variable and the input is the peak wave period and the $10$m wind speed. Thus, the weights in the network are updated optimizing the loss function using only those variables and aims to reproduce the target.

\begin{figure}
    \centering
    \includegraphics[scale=0.7]{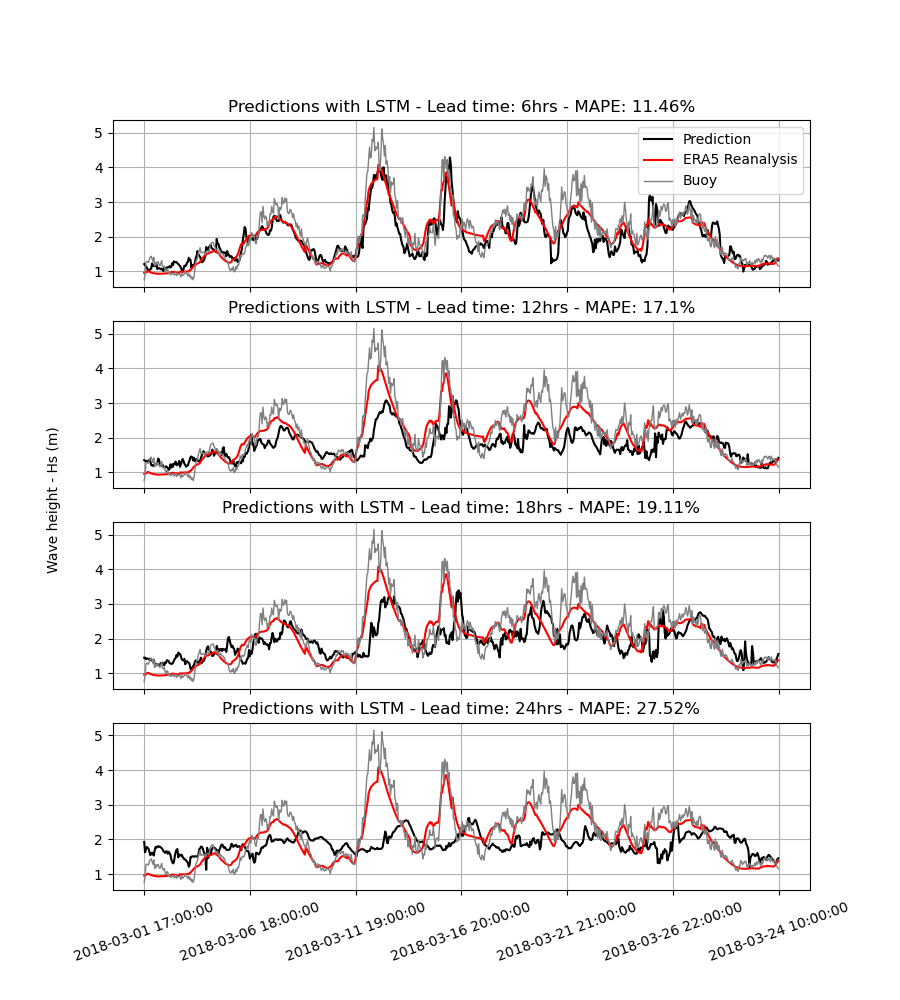}
    \caption{Prediction using LSTM for buoy location number one considering only $T_p$ and $U_{10}$ as the inputs. Predictions are in black, reanalysis from ERA5 in red and buoy data given in grey.}
    \label{fig:15}
\end{figure}

With the same goal of improving the results at location number one, we use the time series for this point in space.  Figure \ref{fig:15} shows the results of this framework of training.  The accuracy drops with this strategy, although are above $70\%$ for all lead times. These results show the obvious dependency of the physical variables (peak period and wind) with the significant height, since the trend of the series is well described for lead time 6. Naturally, the metrics indicate a deterioration in accuracy for larger lead times. The trend in the series is somewhat lost for lead time 24 on, as the peaks  are not reproduced by the LSTM. 

\subsection{Predictions using real buoy data in the training set}

We now present the predictions obtained with buoy measured data as the training set.  Only the significant wave height is considered in the set of features and target. Two buoys locations, namely, number 1 and 2, are used for this analysis. Both locations have buoy data available since April/2009, but there are several missing data and outliers, due to buoy issues and maintenance. Buoy location number two has more erroneous measurements compared with buoy location number one, and therefore the k-Nearest Neighbours classification algorithm was applied to filter these outliers, which allowed the removal of $1.6\%$ of data for location number one and $14.5\%$ for  location number two. A detailed statistics summary can be found in \cite{bose2021assessing}.

\begin{figure}
    \centering
    \includegraphics[scale=0.7]{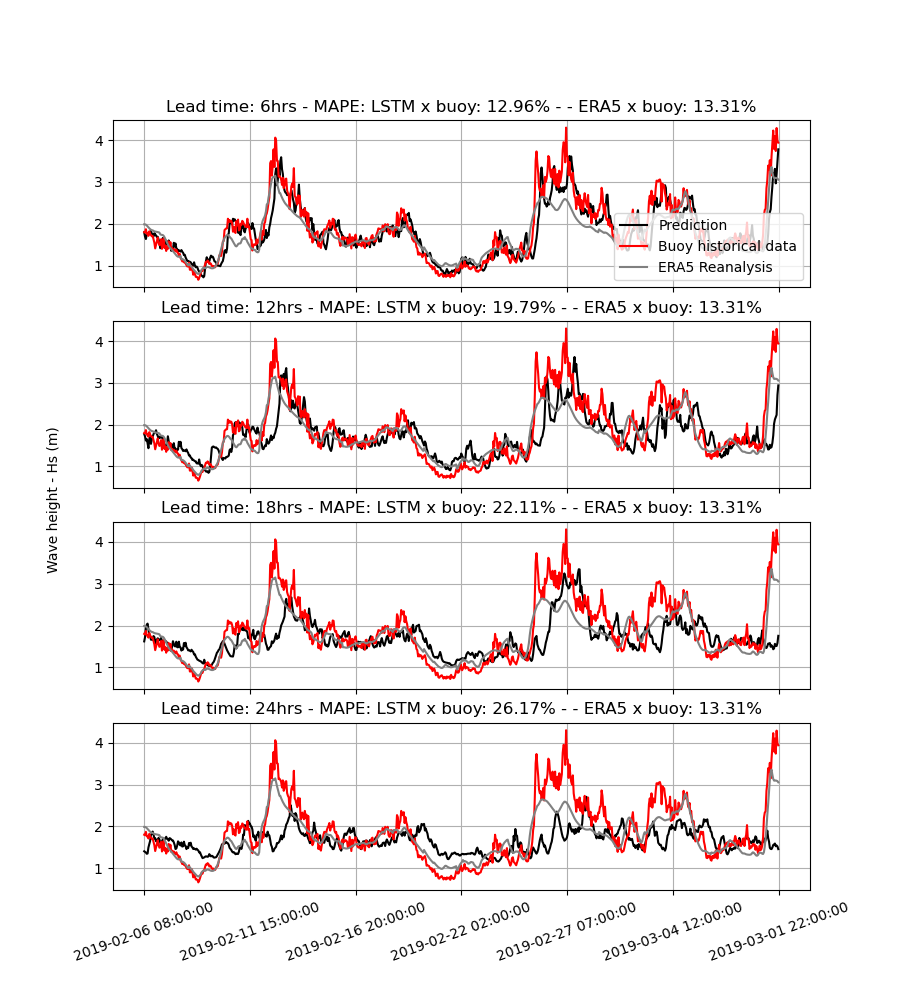}
    \caption{Prediction of $H_s$ using LSTM for buoy location number one using measured data as feature. Predictions are in black and buoy data given in red.}
    \label{fig:16}
\end{figure}

\begin{figure}
    \centering
    \includegraphics[scale=0.7]{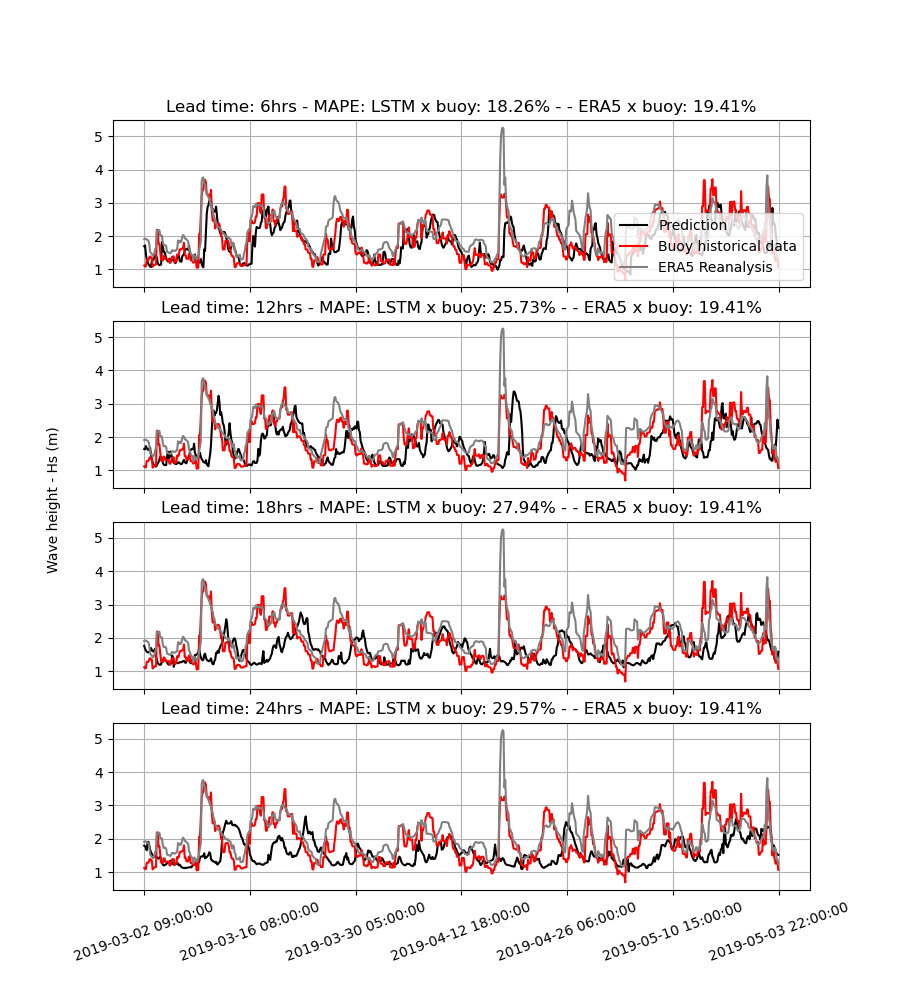}
    \caption{Prediction of $H_s$ using LSTM for buoy location number two using measured data as feature. Predictions are in black and buoy data given in red.}
    \label{fig:17}
\end{figure}

Originally, the data was hourly, but with the outliers removal, gaps in the hourly discretization appeared. Nevertheless, the predictions were made using the same number of points as before, i.e, $751$ steps. The verification time period spans from $06-02-2019$, at $08:00$ GMT to $09-03-2019$, at $17:00$ GMT for buoy location number one and from $02-03-2019$, at $09:00$ GMT, to $30-05-2019$, at $05:00$ GMT for buoy location number two. The same architecture and methodology for training of the LSTM network explained previously was used. Results are presented in Figures~\ref{fig:16} and~\ref{fig:17}. 

It can be observed that the accuracy of the results drop considerably in the predictions using  measured data for the algorithm training, instead of ERA5 data. We obtained the MAPE for those two buoy locations. They are $12.96\%$ ($87.04\%$ accuracy) for lead time 6, $19.79\%$ ($80.21\%$ accuracy) for lead 12, $22.11\%$ ($77.89\%$ accuracy) for lead 18 and $26.17\%$ ($73.83\%$ accuracy) for lead 24 for buoy location number one, and a MAPE of $18.26\%$ ($81.74\%$ accuracy) for lead time 6, $25.73\%$ ($74.27\%$ accuracy) for lead 12, $27.94\%$ ($72.06\%$ accuracy) for lead 18 and $29.57\%$ ($70.43\%$ accuracy) for lead 24 with buoy location number two. The former has a better global accuracy compared to the latter due probablty to the large amount of data that were filtered out, giving a better pattern for the historical time series. Since the removal exclude data points in the historical data, both datasets have gaps in the time domain  and this information could muddle the network. Furthermore, buoy location number one has a pattern that is difficult to predict, compared to buoy location number two, similar behaviour observed with ERA5 data. 

Comparing the accuracy of our proposed methodology with the accuracy of ERA5 data, against the real buoy data, we have a great difference for buoy location number one and a similar result for buoy location number two. Figures~\ref{fig:16} and~\ref{fig:17} show the MAPE metric for ERA5 reanalysis compared to buoy data. Note that ERA5 data is not a forecast, differently from LSTM predictions. For lead time 6h, the LSTM  accuracy is $87.04\%$ against $80.67\%$ of ERA5, in location number one. For location number two, the difference is small, although still superior for LSTM: $81.74\%$ against $80.59\%$.
Considering that ERA5 data is more accurate than any forecast produced with the physical model on whith it is based on, we infer that LSTM with the methodology developed in this work can improve the 6h forecast which uses a physical model.


\section{Conclusion} \label{conclusion}
We have presented a deep learning strategy based on long short-term memory to forecast significant wave height in seven different locations on the Brazil's coast, for four lead times. A three layer architecture is built and training is performed for datasets with different sizes, depending on the lead time.

Results show that accuracy, with respect to the reanalysis data, depending on the location, can reach almost $95\%$ for very short-range forecast. For larger lead times, the accuracy naturally decreases, as the gap between the training set and the prediction becomes larger. Nevertheless, acceptable results are obtained in a computational time that is proportional to the size of the training set. For simulations using only buoy data for training, the accuracy of the LSTM model is $87\%$ for lead time six, and our predictions outperform ERA5 reanalysis results when compared with the observational data. This demonstrates the ability of the methodology presented here  to forecast significant wave height.

The explosion on the number of works and applications which use  machine and/or deep learning algorithms is remarkable. Public domain code libraries and the increasing amount of available datasets (ERA5 has recently increased its historical data back to 1958), appears to be an opportunity to improve present ocean wave forecasting. Despite the very promising advantages of using data-driven models, a better understanding of a number of their operational aspects are necessary, as well as it is needed a stronger mathematical foundation.

Approaches different from the ones used in this work can (and should) be tested. Our strategy was based both on complexity of the network and computational time. Neural networks, although typically faster than physical models, can have its processing time scaled with data size and regions to be modeled. A global prediction, for every grid point in the globe, considering a $0.5$ grid spacing of ERA5, might be not feasible with the method of this paper. In this sense, further studies are necessary to contemplate this goal.

\section*{Acknowledgements}
This work has been funded by the Office of Naval Research Global, under the contract no. N629091812124.

We acknowledge also that this work has been partially funded by the Coordenação de Aperfeiçoamento de Pessoal de Nível Superior - Brasil (CAPES) - Finance Code 001 from the project ROAD-BESM – REGIONAL OCEANIC AND ATMOSPHERIC DOWNSCALING/CAPES, number 88881.146048/2017-01.


 \bibliographystyle{elsarticle-num} 
 \bibliography{bibfile}





\end{document}